# Real-time Characterization of Transient Double-Soliton Dynamics in a Mode-Locked Fiber Laser

**Guomei Wang[1,2], Guangwei Chen[1,2], Wenlei Li[1,2], Chao Zeng[1,2,3] and Wei Zhao[1,2,4]**

[1]*State Key Laboratory of Transient Optics and Photonics, Xi'an Institute of Optics and Precision Mechanics, Chinese Academy of Sciences, Xi'an 710119, China*

[2]*University of Chinese Academy of Sciences (UCAS), Beijing 100049, China*

[3] *e-mail: zengchao@opt.cn*

[4] *e-mail: weiz@opt.ac.cn*

*Abstract*—**By means of the emerging Dispersive Fourier transformation technique, we captured the pulse-resolved spectral evolution dynamics of the double-soliton (DS) states in a single-walled carbon nanotube based Er-doped fiber laser from the initial fluctuations, monitoring the evolution process up to 10 seconds (corresponding to ~260 million roundtrips) discontinuously. Two distinctly different evolutionary types of DS states have been investigated in detail: splitting from one pulse and forming simultaneously. Relaxation oscillations, beating, transient bound state, spectral broadening and pulse interval dynamics have been observed in the evolution process of the DS operation. Our study will be helpful for the further research of mode-locking operation.**

## I. INTRODUCTION

Passively mode-locked fiber lasers have attracted much attention for their wide applications in basic scientific research, biomedical diagnostics, and materials processing [1]-[3]. The passive nonlinear process of the saturable absorber (SA) enables the generation of ultrashort optical pulses in a highly stable train in the fiber lasers [4]. Among the various SAs, single-walled carbon nanotube (SWNT) has been widely utilized due to the advantages of broad operating range, low saturation power, environmental robustness, and easy fabrication [5][6]. Numerous laser modes are locked together to form the ultrashort pulses in the complex mode-locking process [7]. However, passive mode-locked lasers can easily run into the multi-soliton state, especially double-soliton (DS) state [8][9]. Several theories, such as nonlinear cavity feedback, peak power clamping, and gain bandwidth limited pulse splitting, have been proposed to interpret the mechanism of multipulse mode-locking [10]-[12]. The DS state is an extraordinary unwanted phenomenon in some applications of the mode-locked lasers, so, a further research of the evolution of DS is an urgent requirement. This is meaningful for the optimization of mode-locking operation. Nevertheless, the lack of information about how a cavity pulse changes from one round-trip to the next hinders a fundamental understanding of the underlying processes.

Dispersive Fourier transformation (DFT) — also known as real-time Fourier transformation — is an emerging measurement technique which maps the spectrum of an optical pulse to a temporal waveform [13]. The intensity envelope of pulse temporal waveform will mimic the shape of spectrum as long as the pulse propagates inside a dispersive medium sufficiently，which is the temporal equivalent of the far-field condition in diffraction [14]. The DFT technique overcomes the speed limitations of traditional optical instruments and enables fast continuous single-shot measurements in optical sensing, spectroscopy and imaging [15][16]. Based on the DFT technique, Wong *et al.* have revealed the spectral dynamics of dual-color-soliton collisions, multipulse dynamics and the onset of passive mode-locking [17-19]. Buildup dynamics of dissipative soliton and the dissipative rogue waves in ultrafast fiber laser have been investigated by Luo *et al.* using the DFT technique [20][21]. In addition, Zhang *et al.* have demonstrated an ultrafast electrical spectrum analyzer and an real-time broadband radio frequency spectrum analyzer [22][23].

In this paper, we utilized DFT technique as a single-shot spectral-temporal analyzer to investigate the transient DS dynamics in a standard passively mode-locked Er-doped fiber laser (EDFL), which used SWNT film as a mode locker. Two different evolutionary types of DS state have been demonstrated in detail. On the one hand, the two solitons can be formed from pulse splitting. Before the formation of stable double-pulse mode-locking state, pulse interval between the two pulses which split from one pulse increased gradually. On the other hand, both solitons appear simultaneously at the beginning of the mode-locking state. The horizon of DS can be broadened by this result, in addition to previous theoretical and numerical investigations.

## II. EXPERIMENTAL SETUP

The laser system depicted in Fig. 1 consists of a ~ 3.5-m-long erbium-doped fiber (EDF) with 6 dB/m absorption around 980 nm, which is pumped by a 980 nm laser diode (LD) via a polarization-independent tap-isolator-wavelength-division multiplexer (PI-TIWDM). Except the function of a wavelength division multiplexer, the PI-TIWDM is also used to ensure the unidirectional operation and extract intracavity power with a ratio of 10%. Other fiber used in the laser cavity is single mode fiber (SMF) with length of ~4.5 m. SWNT-based SA (SWNT-SA) acts as a mode-locker to initiate the mode-locking operation in the EDFL. A polarization controller (PC) is utilized to optimize the mode-locking performance by adjusting cavity linear birefringence. An opaque metal plate is placed in the free-space section between the LD and PI-TIWDM to control the onset of the EDFL. The ~5-km dispersion-compensating fiber (DCF) serves as a dispersive element for the DFT technique. The output optical power is split to three branches by two optical coupler (OC) for spectral and temporal analyses. The dispersion parameters of EDF, SMF and DCF are approximately −9, 17 and −147 ps/(nm•km) at 1550 nm, respectively. The performances of three branches are measured by a multichannel real-time oscilloscope (6-GHz bandwidth and 20 GSa/s sampling rate) with two high-speed photodetetors (12-GHz bandwidth) and an optical spectrum analyzer (OSA) synchronously. Each channel's record length of the real-time oscilloscope is up to 1 Gpts, which corresponds to a continuous viewing span of 50 ms at a sampling rate of 20 GSa/s. Especially, the real-time oscilloscope used in the experiment possesses some powerful functions, such as 'restraint', 'super segmentation' and 'history', which can restrain the data recording several times with a defined time in a process. This function allows us to monitor a long evolutionary process up to 20 seconds discontinuously, i.e., recording several segments at different time of a process. One branch of the output with ~80% ratio, mapped the optical spectrum of each optical pulse into the time domain through the DFT, is real-time characterized by one channel of the oscilloscope. Other branch of the output with ~10% ratio is monitored directly by another channel of the same real-time oscilloscope. The temporal domain and spectral domain information of the pulses are recorded synchronously by a same trigger signal. Because of the 5 km-DCF, the time delay between the data of two channels is about 24.5 μs.

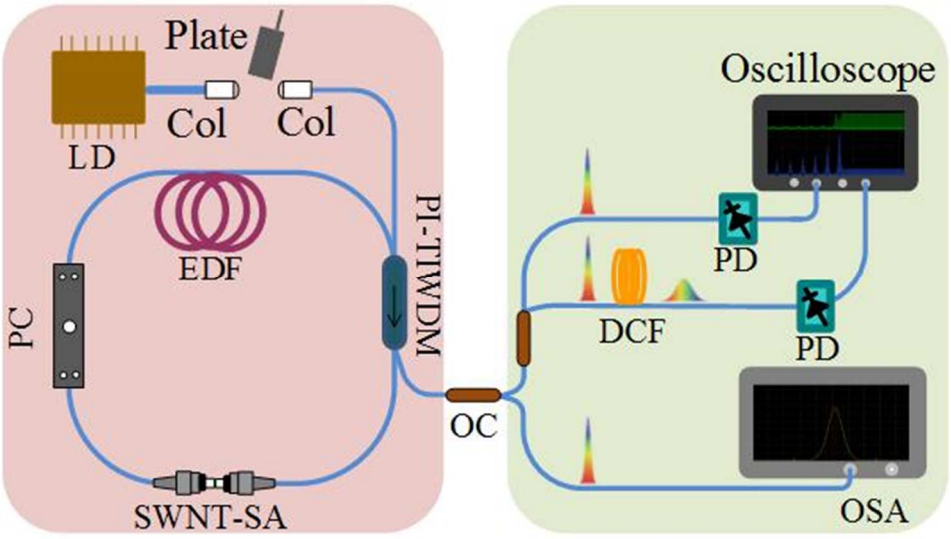

Fig. 1 Experimental setup of the fiber laser and the observation system.

## III. EXPERIMENTAL RESULTS AND ANALYSIS

The double-pulse mode-locking state of the fiber laser is prepared by increasing pump power gradually to 18.5 mW. The repetition rate of the fundamental cavity frequency is ~26 MHz, which corresponds to the total cavity length of ~7.9 m. The DS operation is initiated in a standard approach via opening the opaque metal plate. The evolution process is monitored over a duration of 10 seconds discontinuously, about 25000 consecutive roundtrips (RTs) are recorded in each segment of the data. Constant periodicity is found via the data acquired from the temporal channel and spectral channel, which corresponds to the cavity RT time $T_{rep}$ (38.5 ns) of the fiber laser. To analyze the spectral and spatio-temporal behavior during the evolution process of DS, the time series are segmented into intervals of length $T_{rep}$ to build a two-dimensional evolution diagram. The horizontal axis depicts the dynamics across consecutive RTs and the vertical axis exhibits the information within a single RT.

A spectral evolution dynamic of the DS operation is shown in Fig. 2, which is obtained by the time-wavelength mapping via the insertion of a dispersive element (i.e., DFT technique) [24][25]. Fig. 2(a) reveals the starting dynamics of the mode locking process. Distinct relaxation oscillations are observed before the mode-locking state, which can facilitate the starting of the mode-locking process [26]. The relaxation oscillations occur with period of about 2200 RTs, corresponding to 84.7 μs. Durations of the relaxation oscillations at different evolution process varies from 240 μs to several milliseconds. Obviously, the double-pulse mode-locking begins from single pulse state. Between the relaxation oscillations and the single pulse mode-locking state, the beating dynamics appears two times, as shown in Figs. 2(a) and 2(d). Compared with the first beating dynamics, the second one emerges with slight red-shift. The accumulation of temporal walk-off caused by the Kerr effect over many RTs between pulses results in the beating dynamics [4]. For cognizing the beating dynamics in detail, the variation of amplitude along the dotted line shown in Fig. 2(d) is depicted in the Fig. 2(e). The oscillation frequencies of the two beating dynamics exhibit similar evolution trend: increase gradually and tend to stable, which coincides with the previous report [4]. Pulse splitting emerges soon after the start of mode locking, the mechanism of which has been interpreted by several theoretical attempts [10][12]. After 3 seconds, i.e., ~ 77.9 million RTs, the spectral information of the two pulses in the laser cavity is shown in Fig. 2(b). Because the interval between the two pulses is not large enough, overlaps between the two temporally stretched pulses can be found in the illustration. 4 seconds (i.e., ~103.9 million RTs) after the state shown in Fig. 2(b), the two temporally stretched pulses separate from each other, as shown in the Fig. 2(c). The interval between the two pulses maintains unchanged, and a stable double-pulse mode-locking state is achieved.

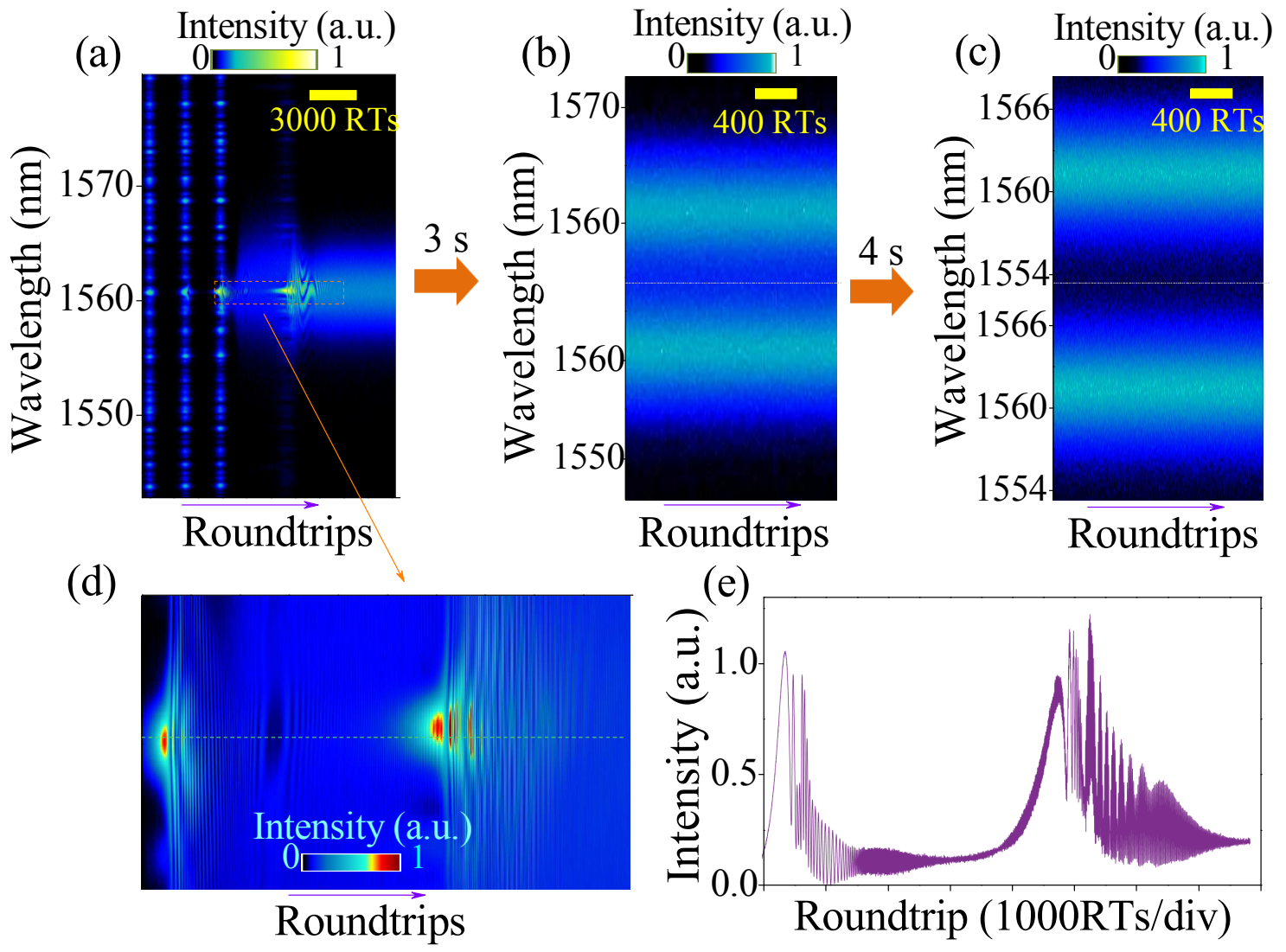

Fig. 2. (a) Starting of the mode locking, (b) and (c) The temporally stretched double-pulse at different time, (d) Close up of the beating dynamics shown in (a), (e) Variation of amplitude along the dotted line shown in (d).

Another spectral evolution dynamic of the DS operation is shown in Fig. 3. Compared to the process shown in Fig. 2, there are some differences between them. Firstly, as shown in Figs. 3(a) and 3(e), a transient bound state is observed at the beginning of the mode locking besides the beating dynamics. Furthermore, we observe that the period from the start of mode-locking to a stable DS state is about 6 seconds (i.e., 155.8 million RTs) here. The single-shot spectrum depicted in Fig. 3(c) reveals the spectral information of each pulse, corresponding to the stable state shown in in Fig. 3(b). It is worthwhile to note that the spectrum of the two solitons owns almost the same appearance. Each spectrum of the both solitons shown in Fig. 3(c) resembles the spectrum of the stable DS state captured by the OSA, which is exhibited in the Fig. 3(d). Remarkably, we are unable to find any previous literature report of the transient bound state in the starting of double-pulse mode-locking. The transient bound state facilitates the broadening of the spectrum, which sustains about 5000 RTs approximately. To study the transient bound state more carefully, a three-dimensional display of that is depicted in Fig. 3(e). A predominant shift of the fringe pattern toward shorter wavelengths can be observed in Fig. 3(e), which corresponds to an decreasing of relative phase [15][26]. As shown in Fig. 3(e), the modulation depth of spectra decreases slowly along with RTs, which means that one of the two bound solitons becomes weaker and weaker [27].

The field autocorrelations of the momentary bound state can be represented by the Fourier transforms of each single-shot spectrum, which traces the separation between two solitons. As sketched in Fig. 3(f), the separation between the two solitons increases rapidly at the initial stage of the transient bound state then remains approximately constant at ~ 18.3 ps for ~ 1500 RTs. Finally, both solitons depart from each other gradually. With the disappearance of one soliton, single pulse mode-locking state appears in the laser cavity. After ~6 s, the stable double-pulse state is formed in the laser cavity through pulse splitting.

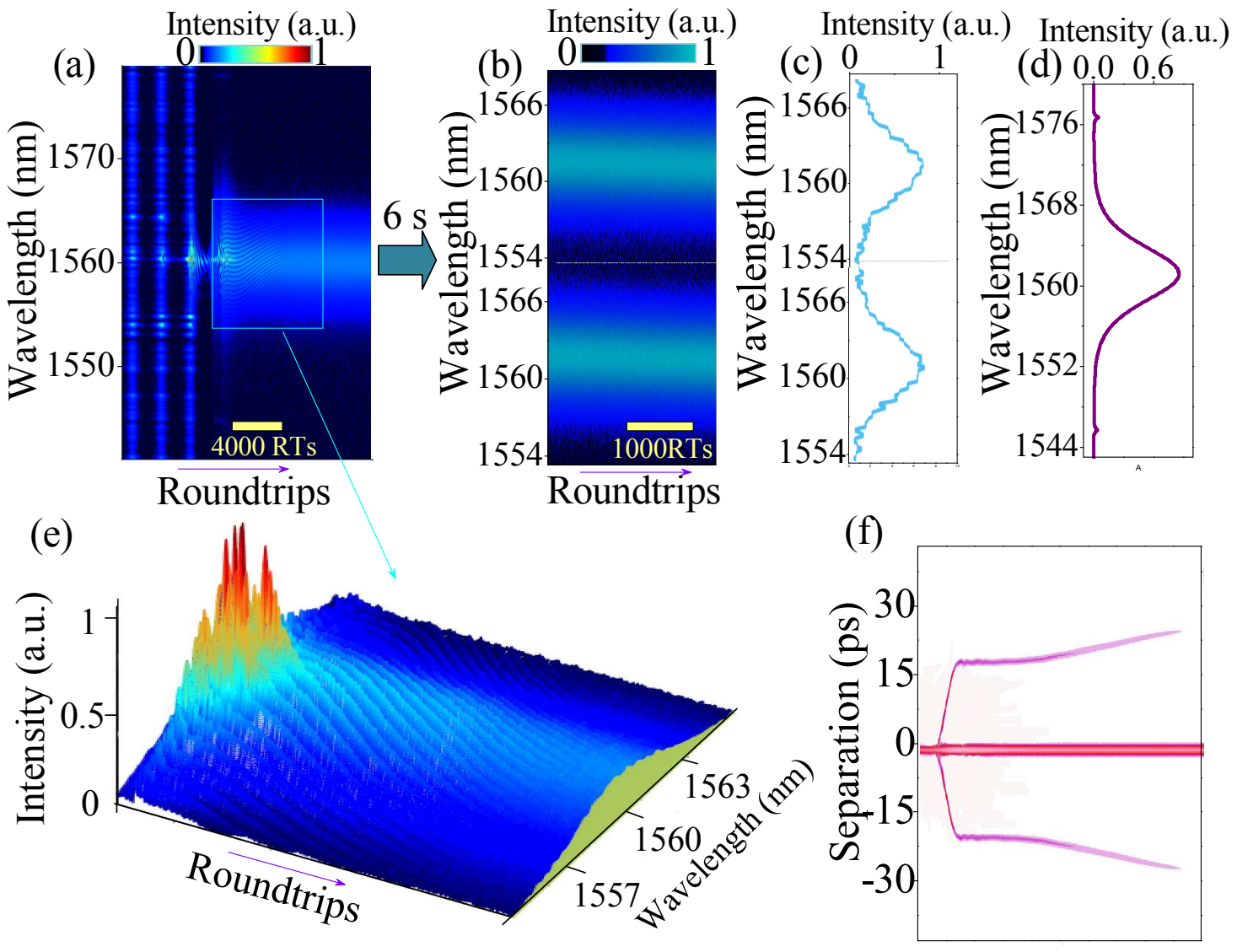

Fig. 3 (a) Starting of the mode locking, (b) and (c) The spectral information of the two solitons, (d) The spectrum recorded by OSA, (e) Three-dimensional display of the transient bound state shown in (a), (f) The separation between both solitons of the bound state.

Both aforementioned DS states evolve from single pulse. Furthermore, we experimentally demonstrate that the two pulses can emerge in the laser cavity almost at the same time through beating fluctuations, respectively, as revealed in Fig. 4. The spectral evolution process of this double-pulse mode-locking and the enlargement of two beating dynamics are shown in Fig. 4(a). Interestingly, the spectral evolutions of these two pulses have some differences in detail during the formation of the mode-locking, such as spectral broadening, spectral shift and beating dynamics. It is also noticed that not all the beating fluctuations can evolve into the broadening of spectrum, as illustrated in Fig. 4(a). The amplitude evolutions of the fluctuations (indicated by arrows in Fig. 4(a)) are shown in Fig. 4(b). As shown in the close-up of Fig. 4(b), the power of the two dominant fluctuations (blue and red lines), which evolve into the mode locked pluses, oscillates more rapidly when the mode-locking is to be formed. Specifically, the oscillation frequencies of them are distinct, which can also be found from the comparison of the two beating dynamics shown in the right of Fig. 4(a). The blue and red lines exhibit identical intensity at the DS state, indicating that the two pulses possess same pulse energy, which actually confirms the energy quantization theory [24]. The fluctuation which does not evolve into mode-locked pulse decays in the laser cavity during the formation of the DS operation, as the yellow line shown in Fig. 4(b). The direct temporal information of the mode-locking process acquired from the undispersed detection of the output train is illustrated in Fig. 4(c). Moreover, after running 4 seconds, the interval between the two pulses is almost unchanged. That indicates a stable DS operation is acquired in the EDFL.

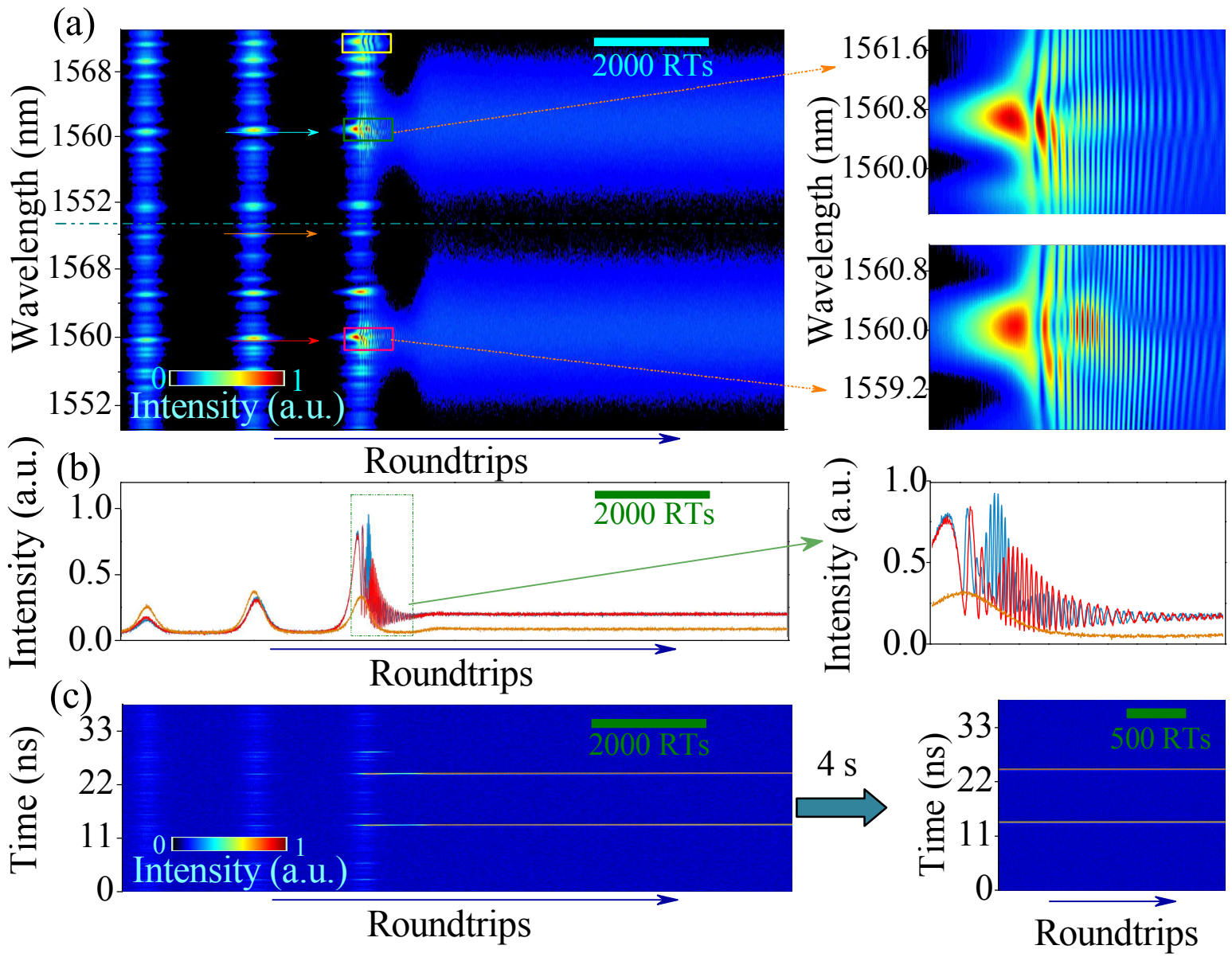

Fig. 4 (a) The starting of the mode locking, and the close-up of the beating dynamics, (b) Normalized power fluctuations along three arrows sketched in (a), (c) Temporal evolution dynamics of the DS.

## IV. Conclusion

In summary, we have studied the evolution dynamics of the DS states in a passively mode-locked EDFL by the DFT technique. Two different evolutionary types of DS state have been investigated in detail: splitting from one pulse and forming simultaneously. Stable DS states can be formed with different evolution time, though they are all formed by pulse splitting. The evolution process of the DS states, including relaxation oscillations, beating, transient bound state, spectral broadening and the pulse interval dynamics, has been observed. To the best of our knowledge, the transient bound state in the starting of the double-pulse mode-locking was observed for the first time. We believe that our works will facilitate a deep understanding of the passively mode-locked all-fiber lasers.

**Funding.** National Natural Science Foundation of China (NSFC) (Grant No. 61635013) and the Strategic Priority Research Program of the Chinese Academy of Sciences (Grant No. XDB24030600).